# Measurement of permeability for ferrous metallic plates using a novel lift-off compensation technique on phase signature

Mingyang Lu[1], Ruochen Huang[1], Wuliang Yin[1], *Senior Member, IEEE*, Qian Zhao[2], Anthony Peyton[1]

[1] School of Electrical and Electronic Engineering,
University of Manchester, Manchester, M13 9PL, UK
[2] College of Engineering, Qufu Normal University, Shandong, China
Corresponding author: wuliang.yin@manchester.ac.uk

*Abstract-* Lift-off of sensor affects the prediction of electromagnetic properties for both ferrous and non-ferrous steel plates. In this paper, we developed a strategy to address this issue for ferrous plates. With increased lift-off, the phase of the measured impedance for steel plates reduces. Meanwhile, the magnitude of the impedance signal decreases. Based on these facts, a phase compensation algorithm is developed which corrects the phase change due to lift-off considering the magnitude of the impedance signal. Further, a new magnetic permeability prediction technique is presented, which has been validated by analytical and measured results. With this new technique, the error in permeability prediction is less than 2% within the range of lift-offs tested.

## I. Introduction

Electromagnetic (EM) technique has been applied for various implications, for example, electrical conductivity perdition, magnetic permeability measurement, surface crack detection, and non-destructive online welding [1]–[6]. Nevertheless, sensors lift-offs can influence the performance of these EM implications. Novel sensor setup, induced responses post-demodulation, and measurement approaches [7]–[9] were used to decrease the error caused by sensors lift-offs. Currently, a few types of research have proposed some new techniques based on the impedance phase feature of the measured multi-frequency spectra to compensate measurements error due to lift-offs [10]. Although the phase can usually be deduced from some analytical approaches such as Dodd Deeds method and finite edge-element technique, most of them commonly require sophisticated and tedious calculations which are inefficient and impossible for some simultaneous testing techniques such as online measurements and welding post-inspection [11]. Consequently, more efficient and compendious methods are imperative to compute the multi-frequency impedance phase, which can be used for various EM applications such as the steels parameters reconstruction and surface crack inspection. Although the proposed approach from [10] was verified to be able to reduce the sensors lift-offs effects on the output multi-frequency impedance phase, the phase error caused by the lift-offs is non-negligible under more precise non-contact measurement with significant lift-offs. Moreover, most of the aforementioned works are related to non-magnetic specimens or just utilize simple characteristics of impedance phase of ferrous specimens.

This paper proposes a novel algorithm to reduce the error of impedance phase for ferrous steels due to sensors lift-offs. The algorithm is based on two basic features. For one side, multi-frequency impedance phases will grow with reduced lift-offs. For the other side, the amplitude of the detected induced response (impedance) will rise up with small sensor lift-offs. Based on this sophisticated phase compensating algorithm, ferrous plate magnetic permeability can be deduced from the measured impedance. Comparing the analytical and measured results for some duplex-phase (DP) steels specimens with various magnetic permeability, this approach has been proved to be accurate enough for the measurement of ferrous plate's magnetic permeability.

## II. EM Sensors Setup

As can be seen from figure 1 and table 1, considering sensors accessibility for experiments and analytical simulations, EM sensor was designed to be 2 co-axially coupled air-cored loop coils: excitation coils and pick-up coils with identical size turns and materials (copper coil). In table 1, a series of lift-off spacers are used to test the lift-off influences on the impedance phase.

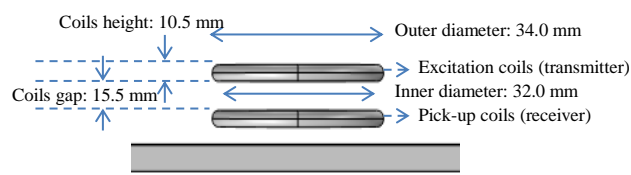

Fig. 1. EM Sensor

Table 1. Probes properties

|  | Values |
|---|---|
| $2r_1$ (Inner diameter)/mm | 32.0 |
| $2r_2$ (Outer diameter)/mm | 34.0 |
| $l_o$ (lift-offs)/mm | 0.8, 2.3, 3.8 |
| $h$ (coils height)/mm | 10.5 |
| $g$ (coils gap)/mm | 15.5 |
| Number of turns $N_1 = N_2$ ($N_1$ - Transmitter; $N_2$ - Receiver) | 30 |

## III. COMPENSATION OF IMPEDANCE PHASE ERROR DUE TO LIFT-OFFS

From our previous researches, the magnitude of the detected response impedance and a frequency feature - zero-crossing frequency were found to grow with reduced sensor lift-offs [12] – [14]. It is also observed that the impedance phase rise up slightly with reduced lift-offs. Consequently, it is speculated that a novel approach could be deduced for compensating the impedance phase error due to sensor lift-offs with the signal amplitude and zero-crossing frequency. The derivations process for compensating the zero-crossing frequency $\omega_0$ was carried out in [13]. Procedure of the proposed algorithm for impedance phase compensation is summarized in figure 2.

For the previous work, the compensated zero-crossing frequency is $\omega_0 = \pi^2 \omega_1/(\pi^2 + 4\ln(\Delta L_0/\Delta L_m))$. Where, $\omega_0$ denotes the zero-crossing frequency after compensation; $\omega_1$ is zero-crossing frequency under current unknown lift-off; $\Delta L_0$ is the inductance amplitude under the high-frequency (when the response signal barely changes with frequencies) with unknown lift-offs; $\Delta L_m$ is the inductance amplitude under same frequencies with the smallest lift-off.

In figure 2, $l_0$ denotes the unknown lift-off; $\theta_r$ denotes the measured phase under any frequency $\omega$ and an unknown lift-off; $\Delta\theta$ denotes the impedance phase change caused by the unknown lift-off, which should be compensated. $\theta$ denotes impedance phase (i.e. $\theta = \theta_r - \Delta\theta$) after compensation.

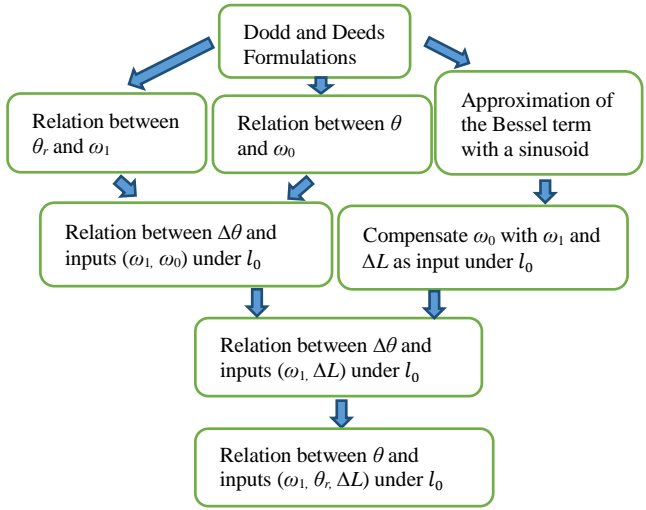

Fig. 2. The procedure of impedance phase compensating deduction

Derivation process of the phase compensation algorithm:

For the metallic plates with $u_r \gg 1$ (ferrous plates), the compensated phase $\varphi(\alpha_0)$ and measured phase $\varphi(\alpha_{0r})$ under unknown lift-off equal,

$$\varphi(\alpha_0) = \frac{1-\sqrt{1/\mu_r^2 + j\omega\sigma\mu_0/\mu_r\alpha_0^2}}{1+\sqrt{1/\mu_r^2 + j\omega\sigma\mu_0/\mu_r\alpha_0^2}} \quad (1)$$

$$\varphi(\alpha_{0r}) = \frac{1-\sqrt{1/\mu_r^2 + j\omega\sigma\mu_0/\mu_r\alpha_{0r}^2}}{1+\sqrt{1/\mu_r^2 + j\omega\sigma\mu_0/\mu_r\alpha_{0r}^2}} \quad (2)$$

Here, $\alpha_0$ is a spatial frequency indicating the geometry feature of the sensor. [6, 12, 13, and 14]

Neglecting $1/\mu_r^2$ term in above equation and assigning $\omega_0 = \mu_r\alpha_0^2/\mu_0\sigma$, $\omega_1 = \mu_r\alpha_{0r}^2/\mu_0\sigma$

The compensated phase $\varphi$ and measured $\varphi$ under unknown lift-off can be expressed as followings,

$$\varphi(\alpha_0) = \frac{1-\sqrt{j\omega/\omega_0}}{1+\sqrt{j\omega/\omega_0}} = \frac{1-\frac{\sqrt{2\omega/\omega_0}}{2}(1+j)}{1+\frac{\sqrt{2\omega/\omega_0}}{2}(1+j)} \quad (3)$$

$$\varphi(\alpha_{0r}) = \frac{1-\sqrt{j\omega/\omega_1}}{1+\sqrt{j\omega/\omega_1}} = \frac{1-\frac{\sqrt{2\omega/\omega_1}}{2}(1+j)}{1+\frac{\sqrt{2\omega/\omega_1}}{2}(1+j)} \quad (4)$$

Then, the measured phase under unknown lift-off should be,

$$\theta_r = \tan^{-1}\left(\frac{\text{Im}(Z_r)}{\text{Re}(Z_r)}\right) = \tan^{-1}\left(\frac{-\text{Re}(L_r)}{\text{Im}(L_r)}\right)$$
$$= \tan^{-1}\left(\frac{-\text{Re}(\phi(\alpha_{0r}))}{\text{Im}(\phi(\alpha_{0r}))}\right) \quad (5)$$
$$= \tan^{-1}\left(\frac{\sqrt{2\omega_1/\omega}}{1-\omega_1/\omega}\right)$$

Similarly, the compensated phase can be derived from $\omega_0$,

$$\theta = \tan^{-1}\left(\frac{\sqrt{2\omega_0/\omega}}{1-\omega_0/\omega}\right) \quad (6)$$

Therefore, the phase change caused by the lift-off should be,

$$\Delta\theta = \theta_r - \theta$$
$$= \tan^{-1}\left(\frac{\sqrt{2\omega_1/\omega}}{1-\omega_1/\omega}\right) - \tan^{-1}\left(\frac{\sqrt{2\omega_0/\omega}}{1-\omega_0/\omega}\right) \quad (7)$$

Then, the compensated phase should be,

$$\theta = \theta_r - \Delta\theta$$
$$= \theta_r - \tan^{-1}\left(\frac{\sqrt{2\omega_1/\omega}}{1-\omega_1/\omega}\right) + \tan^{-1}\left(\frac{\sqrt{2\omega_0/\omega}}{1-\omega_0/\omega}\right) \quad (8)$$

As shown in the appendix, the relation between $\omega_0$ and $\omega_1$ is $\omega_0 = \pi^2\omega_1/(\pi^2 + 4\ln(\Delta L_0/\Delta L_m))$. And the mathematic derivation details of this compensated zero-crossing frequency are shown at the end of the paper.

Finally, the impedance phase after compensation is evaluated from $\omega_1$, $\Delta L_0$, and $\Delta L_m$.

$$\theta = \theta_r - \Delta\theta$$
$$= \theta_r - \tan^{-1}\left(\frac{\sqrt{2\omega_1/\omega}}{1-\omega_1/\omega}\right) \quad (9)$$
$$+ \tan^{-1}\left(\frac{\sqrt{2\omega_1/\left(1+\frac{4}{\pi^2}\ln\left(\frac{\Delta L_0}{\Delta L_m}\right)\right)\omega}}{1-\omega_1/\left(1+\frac{4}{\pi^2}\ln\left(\frac{\Delta L_0}{\Delta L_m}\right)\right)\omega}\right)$$

Assigning $G(\omega) = \tan^{-1}(\sqrt{2\omega_1/\omega}/(1-\omega_1/\omega))$, through some mathematic manipulations, the compensated phase can be obtained.

$$\theta = \theta_r - \Delta\theta$$
$$= \theta_r - G(\omega) + G\left(\left(1+\frac{4}{\pi^2}\ln\left(\frac{\Delta L_0}{\Delta L_m}\right)\right)\omega\right) \quad (10)$$

With, $$G(\omega) = \tan^{-1}\left(\frac{\sqrt{2\omega_1/\omega}}{1-\omega_1/\omega}\right) \quad (11)$$

Where, $\Delta L_0$ is the inductance amplitude under the high - frequency (when the response signal barely changes with frequencies) with unknown lift-offs; while $\Delta L_m$ is the inductance amplitude under same frequencies with the smallest lift-off (here this lift-off in measurement setup is 0.8 mm).

It can be seen in equation 11 that with the measured phase, inductance magnitude and zero-crossing frequencies from the measurements at an unknown lift-off as inputs, impedance phases $\theta$ after compensating (phase with zero lift-offs) could be obtained using the compensation scheme proposed above. For instance, if the sensor is put on a lift-off approaching 0, $\ln(\Delta L_0/\Delta L_m)$ should equal 0. As a result, the corresponding compensated result $\theta_0$ calculated from equation 10 equals $\theta_r$, which is reasonable under a negligible lift-off.

## IV. ANALYTICAL SOLUTIONS AND MEASUREMENTS

In order to validate the feasibility of the deduced phase compensating approach, measurements and analytical solution have been made to compare impedance phases with various sensor lift-offs. The co-relation between impedance and inductance are shown in followings:

$$\Delta L = \frac{Z - Z_{air}}{j\omega} \quad (12)$$
$$\omega = 2\pi f \quad (13)$$

Here, $Z$ represents the sensor mutual impedance with specimens; $Z_{air}$ denotes the mutual impedance between the sensor's transmitter and receiver without specimens; $f$ is the operation frequency.

Then, the impedance phase can be evaluated,

$$\theta = \tan^{-1}\left(\frac{\text{Im}(Z-Z_{air})}{\text{Re}(Z-Z_{air})}\right) = \tan^{-1}\left(\frac{-\text{Re}(\Delta L)}{\text{Im}(\Delta L)}\right) \quad (14)$$

### A. Analytical solutions

For the analytical solutions, Dodd Deeds approach [11] was utilized to compute the sensor's detected response signal - impedance. The sample was chosen to be a duplex-phase specimen - DP600 (specimen's properties and size data are shown in Table 2) under varying lift-offs of 0.8 mm, 2.3 mm, and 3.8mm. The analytical solver is scripted and operated on MATLAB coding platform, which is utilized for the evaluation of inductance $\Delta L$ (equations 15 - 20 in the appendix) and the compensated phase using equation 10.

### B. Measurements

In order to measure the impedance/inductance phase of the samples, a symmetric air-cored electromagnetic sensor was designed for steel micro-structure monitoring in the Continuous Annealing & Processing Line (CAPL). As can be seen from Fig. 3, the excitation coil sits in the middle and two receive coils at bottom and top respectively. The geometry profile of the sensor is illustrated in Table 1. Receive coil 2 is used as the test coils; receive coil 1 is served as a reference coil. In the paper, only receive 2 coil signal is recorded and served as the response output signal. All the coils have the same diameters, i.e. an inner diameter of 32.0 mm and an outer diameter of 34.0 mm. Each of the coils has 30 turns, and the coil separation is 35.0 mm. SI 1260 impedance analyser has been utilized to measure the air-core sensor induced signal response – mutual impedance or inductance of the sensor influenced by the tested samples. The working frequency range of the instrument is set from 310 Hz to 3 MHz. Moreover, all the samples are tested under a series lift-offs of 0.8, 2.3, and 3.8 mm.

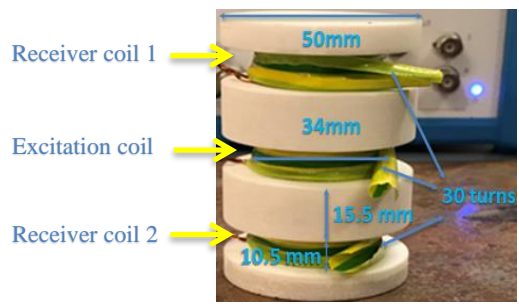

(a)

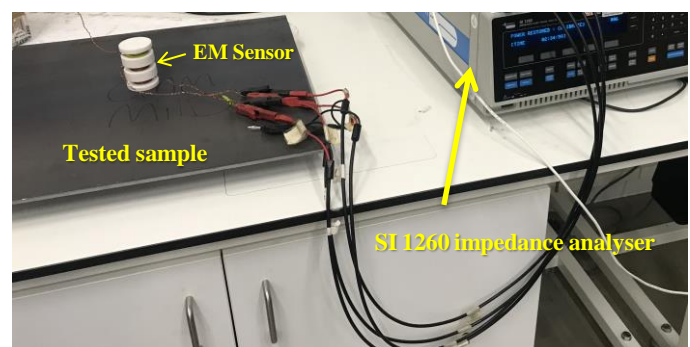

(b)

Fig. 3. Measurement setup a) EM air-cored Sensor configuration b) SI 1260 impedance analyser

### C. Results

Figure 4 exhibits both the real part and imaginary part of the simulations and measurements of sensor-plates system mutual inductance multi-frequency spectra. In figure 4, it is obviously that inductance curves magnitude drop off with increased lift-offs. Meanwhile, the zero-crossing frequency decreases with increased lift-offs. Some singular points may be encountered during the measurements which are due to the signal noise of SI 1260 impedance analyser, especially under the low frequency.

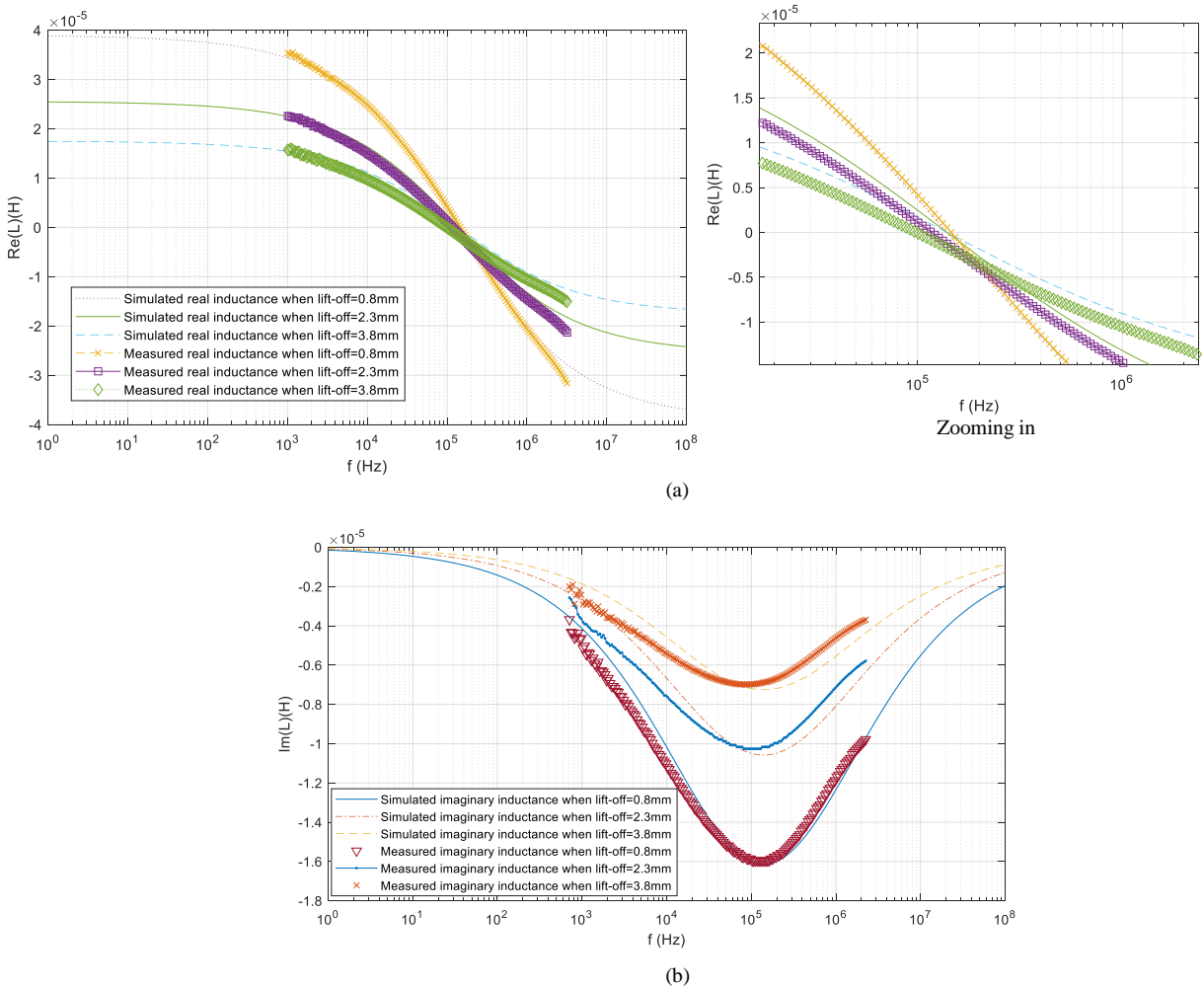

Fig. 4. Real and imaginary part of inductance under varying lift-offs - 0.8 mm, 2.3 mm, and 3.8mm (a) real part (b) imaginary part

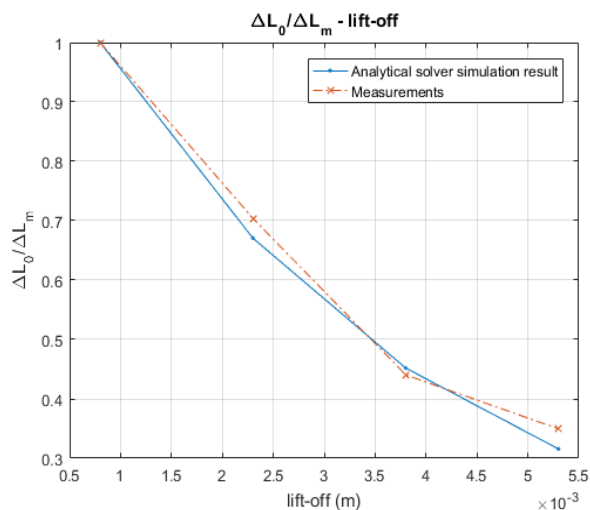

Fig. 5. Trend of inductance term $\Delta L_0/\Delta L_m$ (for DP 600 specimen) for different lift-offs

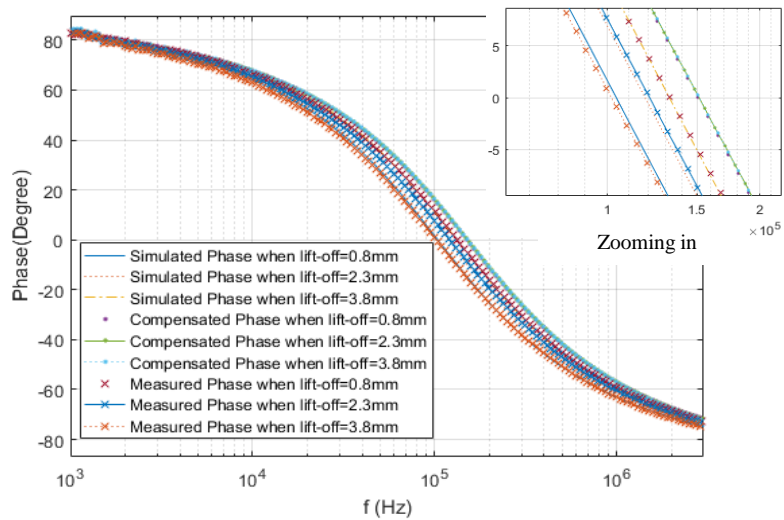

Fig. 6. Compensation performance both on simulations and measurements with 0.8, 2.3, 3.8 mm lift-offs

Table 2. Properties of Duplex-phase specimens

| Specimens | Electrical conductivity (MS/m) | Relative permeability | Planar size (mm) | Thickness (mm) |
|---|---|---|---|---|
| DP600 | 4.13 | 222 | 500 × 400 | 7.0 |
| DP800 | 3.81 | 144 | 500 × 400 | 7.0 |
| DP1000 | 3.80 | 122 | 500 × 400 | 7.0 |

Table 3. Relative permeability measurements for different lift-offs

| Plates | Lift-offs (mm) | Actual relative permeability | Relative permeability without compensation | Relative permeability inferred from compensated phase | Relative error for non-compensated permeability | Relative error for compensated permeability |
|---|---|---|---|---|---|---|
| DP800 | 0.8 | 144 | 138.48 | 142.37 | 3.83% | 1.13% |
|  | 2.3 | 144 | 136.76 | 142.12 | 5.03% | 1.31% |
|  | 3.8 | 144 | 133.31 | 141.94 | 7.42% | 1.43% |
| DP1000 | 0.8 | 122 | 117.68 | 120.72 | 3.54% | 1.05% |
|  | 2.3 | 122 | 115.27 | 120.57 | 5.52% | 1.17% |
|  | 3.8 | 122 | 111.98 | 120.15 | 8.21% | 1.52% |

In figure 5, it is observed that the inductance term $\Delta L_0/\Delta L_m$ decreases with increased lift-offs. Consequently, the relative loss of the inductance $\Delta L_0/\Delta L_m$ can be used for the compensation of inductance or impedance phase due to lift-offs. which can be used to compensate the drop in phase with rising lift-offs. Here, $\Delta L_m$ is the sample's inductance (as shown in equation 12) with end point frequency (the last frequency sample for both simulations and measurements) for the smallest lift-off (0.8 mm under the sensor setup in figure 3).

Figure 6 shows the simulations, measurements, and the phase multi-frequency spectra after the proposed compensation algorithm (equation 10 and 11). It can be seen that both the simulated and measured phase decrease as increased lift-offs. In addition, the compensated phase is barely affected by the lift-off. Based on the compensated phase, ferrous plate magnetic permeability could be easily predicted via the measured response of the sensor. The ferrous metallic plates' magnetic permeability measurement technique is validated via the comparison of modelling and measured data for the mentioned sensor next to dual-phase (DP) steels with various values of magnetic permeability.

In principle, the magnetic permeability reconstruction for the tested specimens is finding the simulated multi-frequency inductance/impedance curve (via equations 15 - 20 in appendix) that is closest to the measured multi-frequency spectra data (after the proposed compensation algorithms - equation 10 and 11) while changing the permeability. In order to validate the proposed magnetic permeability reconstruction technique via the compensated phase, multi-frequency inductances of two ferrous specimens have been tested (specimens' properties and size data are shown in Table 2). In the measurement process, 120 logarithmically spaced frequencies samples range from 310 Hz to 3 MHz have been chosen as the operation frequencies. In addition, both DP steel specimens have identical size of 500 × 400 × 7.0 mm. Consequently, magnetic permeability comparisons for compensated phase and the measured phase without compensation is shown in table 3.

It can be concluded from table 3 that the magnetic permeability reconstruction shows a better performance through the proposed impedance or inductance phase compensation scheme (equation 10 and 11).

In practical application, the lift-offs range may be different. However, it has been found that the error of the measured permeability is always within a small value of 5%.

## V. Conclusions

In this paper, a compensation technique is developed for the relief of lift-off effects on impedance phase for metallic ferrous plates. From the results, it can be easily observed that both phase and signal (impedance/inductance) magnitude decreases as lift-off increases. And the measured inductance or impedance can be used for the compensation of impedance phase loss due to lift-offs via the proposed algorithms. Based on the proposed phase compensation approach, a magnetic permeability measurement technique was proposed that is also virtually independent of lift-offs. The results have been verified with both measurements and simulations of selected cases.

## Acknowledgments

This work was supported by [UK Engineering and Physical Sciences Research Council (EPSRC)] [grant number: EP/M020835/1] [title: Electromagnetic tensor imaging for in-process welding inspection].

## APPENDIX

Inferring the compensated zero-crossing frequency (ZCF) $\omega_0$ from the measured ZCF and inductance ($\omega_1$ and $\Delta L$) under a lift-off of $l_0$:

Dodd Deeds method has been chosen as the forward simulation solver for the calculation of the inductance due to the appearance of the sample when tested by an axially symmetric air-cored sensor [15].

The inductance due to the appearance of the sample is the subtraction of the sensor tested inductance when sensor is put on a specimen ($L(\omega)$) and that when sensor is in empty region ($L_A(\omega)$): $\Delta L(\omega) = L(\omega) - L_A(\omega)$.

Dodd Deeds formulations are listed as follows:

$$\Delta L(\omega) = K \int_0^\infty \frac{P^2(\alpha)}{\alpha^6} A(\alpha)\phi(\alpha)d\alpha \quad (15)$$

Where,

$$A(\alpha) = (1 - e^{-2\alpha h})e^{-\alpha(G+h+2l_0)} \quad (16)$$

$$\phi(\alpha) = \frac{(\mu_r \alpha - \alpha_1)}{(\mu_r \alpha + \alpha_1)} = \frac{\mu_r \alpha - \sqrt{\alpha^2 + j\omega\sigma\mu_r\mu_0}}{\mu_r \alpha + \sqrt{\alpha^2 + j\omega\sigma\mu_r\mu_0}}$$
$$= \frac{1 - \sqrt{1/\mu_r^2 + j\omega\sigma\mu_0/\mu_r\alpha^2}}{1 + \sqrt{1/\mu_r^2 + j\omega\sigma\mu_0/\mu_r\alpha^2}} \quad (17)$$

$$K = \frac{\pi\mu_0 N^2}{h^2(r_1 - r_2)^2} \quad (18)$$

$$P(\alpha) = \int_{\alpha r_1}^{\alpha r_2} xJ_1(x)dx \quad (19)$$

$$\alpha_1 = \sqrt{\alpha^2 + j\omega\sigma\mu_r\mu_0} \quad (20)$$

$l_0$ is sensor's lift-off; $h$ is sensor's coil height; $N$ is sensor's coil turn number; $r_1$ and $r_2$ are inner and outer radii of sensor's coil; $\mu_r$ is the specimen's relative permeability. $\mu_0$ is the vacuum permeability; $G$ is the distance between the excitation coil and receiving coil.

In equation (15), since the $\phi(\alpha)$ term barely change with $\alpha$ (compared with $A(\alpha)$ and $P(\alpha)$), $\phi$ could be estimated as equation (22),

$$\Delta L(\omega) = \phi(\alpha_0)\Delta L_0 \qquad (21)$$

$\alpha_0$ is the spatial frequency, which is a constant controlled by sensor configuration.

From equation (21), the phase of tested inductance or impedance is merely controlled via $\phi(\alpha_0)$.

Where, $\phi(\alpha_0) = \dfrac{-\sqrt{1/\mu_r^2 + j(\mu_0/\mu_r)\sigma\omega\alpha_0^2} + 1}{\sqrt{1/\mu_r^2 + j(\mu_0/\mu_r)\sigma\omega\alpha_0^2} + 1} \qquad (22)$

Neglect $1/\mu_r^2$ in equation (22),

$$\phi(\alpha_0) = \dfrac{-\sqrt{j(\mu_0/\mu_r)\sigma\omega\alpha_0^2} + 1}{\sqrt{j(\mu_0/\mu_r)\sigma\omega\alpha_0^2} + 1} \qquad (23)$$

In equation (23), it can be observed that $\phi(\alpha_0)$ is sample and sensor related (controlled by $\sigma$ and $\mu_r$).

Assign $\dfrac{\mu_r\alpha_0^2}{\mu_0\sigma}$ with $\omega_1$, equation (23) can be expressed as,

$$\phi(\alpha_0) = \dfrac{-\sqrt{j\omega/\omega_1} + 1}{\sqrt{j\omega/\omega_1} + 1} \qquad (24)$$

In equation (21), $\Delta L_0$ denotes the magnitude of the tested inductance, which is solely controlled by the sensor configuration (cannot affected by the specimen properties).

From our previously work, a simple function $sin^2\left(\dfrac{\alpha\pi}{2\alpha_0}\right)$ with its maximum at $\alpha_0$ is used to approximate $\Delta L_0$ [13],

$$\Delta L_0 \approx \Delta L_m e^{-2\alpha l_0} \sin^2(\dfrac{\alpha\pi}{2\alpha_0}) \qquad (25)$$

Where $\Delta L_m$ is the sample's inductance (as shown in equation 12) with start point frequency (the first frequency sample for both simulations and measurements) for the smallest lift-off (0.8 mm under the sensor setup in figure 3)

The revised $\alpha$ should maximize $e^{-2\alpha l_0}sin^2\left(\dfrac{\alpha\pi}{2\alpha_0}\right)$ and therefore $e^{-\alpha l_0}sin\left(\dfrac{\alpha\pi}{2\alpha_0}\right)$.

In our previous work [13], the shift in $\alpha_0$ caused by lift-off effect - $\alpha_{0r}$ can be derived as,

$$\alpha_{0r} = \alpha_0 - \dfrac{4\alpha_0^2 l_0}{\pi^2} \qquad (26)$$

Therefore, the revised $\omega_1$ becomes

$$\omega_1 = \dfrac{\left(\alpha_0^2\pi^4 - 8\pi^2\alpha_0^3 l_0 + 16\alpha_0^4 l_0^2\right)\mu_r}{\pi^4\sigma\mu_0} \qquad (27)$$

Combining (25) with (26), $\Delta L_0$ becomes

$$\Delta L_0 = \Delta L_m e^{-2(\alpha_0 - \tfrac{4\alpha_0^2 l_0}{\pi^2})l_0} cos^2(\dfrac{2\alpha_0 l_0}{\pi})$$

$$= \Delta L_m e^{-2(\alpha_0 - \tfrac{4\alpha_0^2 l_0}{\pi^2})l_0} \left(\dfrac{cos(\tfrac{4\alpha_0 l_0}{\pi}) + 1}{2}\right)$$

Considering $\alpha_0 l_0 \ll 1$ and based on small-angle approximation $cos(\theta) \approx 1 - \theta^2/2$, $cos(4\alpha_0 l_0/\pi)$ is substituted with $1 - (4\alpha_0 l_0/\pi)^2/2$.

$\Delta L_0$ becomes, $\Delta L_0 = \Delta L_m e^{-2(\alpha_0 - \tfrac{4\alpha_0^2 l_0}{\pi^2})l_0}(1 - \dfrac{4\alpha_0^2 l_0^2}{\pi^2})$

Substituting $\left(1 - \dfrac{4\alpha_0^2 l_0^2}{\pi^2}\right)$ with $e^{-\tfrac{4\alpha_0^2 l_0^2}{\pi^2}}$

$$\Delta L_0 = \Delta L_m e^{-2(\alpha_0 - \tfrac{4\alpha_0^2 l_0}{\pi^2})l_0} e^{-\tfrac{4\alpha_0^2 l_0^2}{\pi^2}} \qquad (28)$$

$$= \Delta L_m e^{-2(\alpha_0 - \tfrac{2\alpha_0^2 l_0}{\pi^2})l_0}$$

Then,

$$\ln\dfrac{\Delta L_0}{\Delta L_m} = -2(\alpha_0 - \dfrac{2\alpha_0^2 l_0}{\pi^2})l_0 \qquad (29)$$

And further derivation from (29):

$$4\alpha_0^2 l_0^2 - 2\pi^2\alpha_0 l_0 - \pi^2 \ln\dfrac{\Delta L_0}{\Delta L_m} = 0 \qquad (30)$$

This is now a quadratic equation with $\alpha_0 l_0$ as its variable.

Therefore, the solution for $\alpha_0 l_0$ is

$$\alpha_0 l_0 = \dfrac{\pi^2 - \sqrt{\pi^4 + 4\pi^2\ln\tfrac{\Delta L_0}{\Delta L m}}}{4} \qquad (31)$$

Since $\alpha_0 l_0 \ll 1$, the other solution, the other solution $\alpha_0 l_0 = \dfrac{\pi^2 + \sqrt{\pi^4 + 4\pi^2\ln\tfrac{\Delta L_0}{\Delta L m}}}{4}$ therefore is discarded.

From equation (31), lift-off can be estimated as

$$l_0 = \dfrac{\pi^2 + \sqrt{\pi^4 + 4\pi^2\ln\tfrac{\Delta L_0}{\Delta L m}}}{4\alpha_0} \qquad (32)$$

Combining (27) with (32),

$$\omega_1 = \dfrac{\alpha_0^2\left(\pi^2 + 4\ln\tfrac{\Delta L_0}{\Delta L m}\right)\mu_r}{\pi^2\sigma\mu_0} \qquad (33)$$

Further derivation from equation (33),

$$\alpha_0^2(\pi^2 + 4\ln\frac{\Delta L_0}{\Delta L_m})\mu_r - \pi^2\sigma\mu_0\omega_1 = 0$$

And the solution is

$$\alpha_0 = \sqrt{\frac{\pi^2\sigma\mu_0\omega_1}{\left(\pi^2 + 4\ln\frac{\Delta L_0}{\Delta L_m}\right)\mu_r}} \quad (34)$$

Thus, the zero-crossing frequency can be compensated as following,

$$\omega_0 = \frac{\mu_r\alpha_0^2}{\mu_0\sigma} = \frac{\pi^2\omega_1}{\left(\pi^2 + 4\ln\frac{\Delta L_0}{\Delta L_m}\right)} \quad (35)$$